\documentclass[seceq]{ptptex}

\title{Collisions of Shock Waves in General Relativity}
\author{C. \textsc{Barrab\`es}$^{1,}$ and P. A. \textsc{Hogan}$^{2,}$}
\inst{$^1$\small Laboratoire de Math\'ematiques et Physique Th\'eorique,\\
\small Universit\'e F. Rabelais, F\'ed\'eration Denis Poisson-CNRS, \\
\small Parc de GrandMont, 37200 Tours, France\\
$^2$\small School of Physics,\small University College Dublin,
Belfield, Dublin 4, Ireland}
\recdate{}

\abst{We show that the Nariai--Bertotti Petrov type D, homogeneous solution of Einstein's vacuum field equations with a 
cosmological constant describes the space--time in the interaction region following the head--on collision 
of two homogeneous, plane gravitational shock waves each initially traveling in a vacuum containing no cosmological 
constant. A shock wave in this context has a step function profile 
in contrast to an impulsive wave which has a delta function profile. Following the collision two light--like signals, each 
composed of a plane, homogeneous light--like shell of matter and a plane, homogeneous impulsive gravitational wave, travel away from each other 
and a cosmological constant is generated in the interaction region. Furthermore a plane, light--like signal consisting of an electromagnetic shock 
wave accompanying a gravitational shock wave is described with the help of two real parameters, 
one for each wave. The head--on collision of two such light--like signals is examined and we show that if a simple algebraic relation is 
satisfied between the two pairs of parameters associated with each incoming light--like 
signal then the space--time in the interaction region following the collision is a Bertotti space--time which is a 
homogeneous solution of the vacuum Einstein--Maxwell field equations with a 
cosmological constant.}
 
\begin{document}
 \maketitle 
%\end{abstract}       
%\thispagestyle{empty}
%\newpage

\section{Introduction}\indent
Perhaps the best known 4--dimensional pseudo--Riemannian space--time manifold which is a Cartesian product of two 2--dimensional 
manifolds of constant Gaussian curvature is the Bertotti--Robinson \cite{B}\cite{R} solution of the vacuum 
Einstein--Maxwell field equations. The line--element is the sum of the two 2--dimensional line--elements corresponding to each manifold 
in the Cartesian product. It is well known that the Bertotti--Robinson solution of the vacuum Einstein--Maxwell field equations can 
be transformed into the Bell--Szekeres \cite{BS} solution describing the space--time in the interaction region following the 
head--on collision of two homogeneous, plane \emph{electromagnetic} shock waves (see \cite{ES} p.399). As a special case, in the absence of 
electromagnetic fields, Bertotti \cite{B} has given a homogeneous solution of Einstein's vacuum field equations with a cosmological 
constant. This solution was discovered earlier by Nariai \cite{Na} (see also \cite{ES}). Its importance for this paper is 
its association with the Bertotti--Robinson solution and since this association is due to Bertotti we will refer to the solution as the Nariai--Bertotti solution.
 In this paper we show that the Nariai--Bertotti solution describes the space--time in the interaction region following the head--on collision 
of two homogeneous plane \emph{gravitational} shock waves. In addition we study the head--on collision of two homogeneous, plane, light--like signals each consisting of an electromagnetic shock wave 
accompanied by a gravitational shock wave. We show that if a simple relation is satisfied between the two pairs of parameters associated with each incoming light--like 
signal then the space--time in the interaction region following the collision is a Bertotti \cite{B} space--time.

The outline of this paper is as follows: In section 2 the gravitational shock wave collision space--time is constructed and 
interpreted physically in section 3 in which the region of the space--time coinciding with the Nariai--Bertotti space--time is also identified.
In section 4 the space--time model of a gravitational shock wave accompanied by an electromagnetic shock wave is described. This is followed 
in section 5 by the space--times describing the collision of two such light--like signals. Properties of these space--times including the identification of 
the subregion in each case which is a Bertotti space--time, are discussed in section 6.

\setcounter{equation}{0}
\section{Gravitational Shock Wave Collision}\indent
The Nariai--Bertotti \cite{B}\cite{Na} homogeneous solution of Einstein's vacuum field equations with a cosmological constant is the Cartesian product of 
two 2--dimensional manifolds of equal constant curvature, has a line--element which is the sum of the two 2--dimensional line--elements and 
the equal constant curvatures can be positive or negative. In local coordinates $\{x^\mu\}$, 
with $\mu =1, 2, 3, 4$, it is a solution of 
\begin{equation}\label{3.1}
R_{\mu\nu}=\Lambda\,g_{\mu\nu}\ ,\end{equation}
where $g_{\mu\nu}$ are the metric tensor components of the 4--dimensional manifold, $R_{\mu\nu}$ are the components of the Ricci tensor and $\Lambda$ 
is the cosmological constant. We must 
choose a sign for the cosmological constant and we begin by taking $\Lambda <0$. Different representations of the Nariai--Bertotti solution are, of course, possible. In addition 
to those given in \cite{B} and \cite{Na} there is the representation given by Ozsv\'ath \cite{O} in which the solution is derived as a member of the set of all homogeneous 
solutions of Einstein's vacuum field equations with a cosmological constant. A convenient representation for our present purposes however is given by the 
line--element
\begin{equation}\label{3.2}
ds^2=ds'^2+ds''^2\ ,\end{equation}with
\begin{equation}\label{3.3}
ds'^2=d\xi ^2-\cos ^2\sqrt{-\Lambda}\,\xi\,dx^2\ \ {\rm and}\ \ ds''^2=-d\lambda ^2-\cosh ^2\sqrt{-\Lambda}\,\lambda\,dy^2\ .\end{equation}
Labelling the coordinates $\{x^A\}$ with $A=1, 2$ in either case, the Riemann tensor components for each of these manifolds take the form
\begin{equation}\label{3.4}
R'_{ABCD}=-\Lambda\,(g'_{AC}g'_{BD}-g'_{AD}g'_{BC})\ \ {\rm and}\ \  R''_{ABCD}=-\Lambda\,(g''_{AC}g''_{BD}-g''_{AD}g''_{BC})\ ,\end{equation}
respectively, confirming that each 2--dimensional manifold has the same constant curvature. 
In the line--element obtained by substituting (\ref{3.3}) into (\ref{3.2}) we write $\Lambda =-2ab$, where $a$ and $b$ are constants, and define new coordinates $u$ and $v$ 
in place of $\xi$ and $\lambda$ via the transformations
\begin{eqnarray}
\sqrt{-\Lambda}\,\xi&=&au+bv\ ,\label{3.5}\\
\sqrt{-\Lambda}\,\lambda &=&au-bv\ .\label{3.6}\end{eqnarray}This results in (\ref{3.2}) taking the form
\begin{equation}\label{3.7}
ds^2=-\cos ^2(au+bv)\,dx^2-\cosh ^2(au-bv)\,dy^2+2\,du\,dv\ .\end{equation} For the case $\Lambda >0$ we replace (\ref{3.3}) by the line--elements
\begin{equation}\label{3.7'}
ds'^2=-d\xi ^2-\cos ^2\sqrt{\Lambda}\,\xi\,dx^2\ \ {\rm and}\ \ ds''^2=d\lambda ^2-\cosh ^2\sqrt{\Lambda}\,\lambda\,dy^2\ .\end{equation}Now (\ref{3.4}) 
is formally unchanged so that in this case the two 2--dimensional manifolds have equal constant curvatures but of the opposite sign to the equal 
constant curvatures of the manifolds with line--elements (\ref{3.3}). The transformations (\ref{3.5}) and (\ref{3.6}) are now replaced by $\sqrt{\Lambda}\,\xi
=au+bv$ and $\sqrt{\Lambda}\,\lambda =au-bv$ and these lead once again to (\ref{3.7}) from (\ref{3.2}) and (\ref{3.7'}).

In the space--time with line--element (\ref{3.7}) there are two families of intersecting 
null hyperplanes, $u={\rm constant}$ and $v={\rm constant}$. This space--time describes the gravitational field in the interaction region $u>0, v>0$ following 
the head--on collision of two gravitational shock waves. To see this we replace $u$ and $v$ in (\ref{3.7}) by $u_+=u\vartheta (u)$ and $v_+=v\vartheta (v)$ 
respectively, where $\vartheta (u)$ is the Heaviside step function which is equal to unity if $u>0$ and vanishes if $u<0$ (similarly for $\vartheta (v)$). Now 
(\ref{3.7}) is replaced by
\begin{equation}\label{3.8}
ds^2=-(\theta ^1)^2-(\theta ^2)^2+2\,\theta ^3\,\theta ^4=g_{ab}\theta ^a\,\theta ^b\ ,\end{equation}with the 1--forms $\{\theta ^a\}$, with $a=1, 2, 3, 4$ given by
 \begin{equation}\label{3.9}
 \theta ^1=\cos (au_++bv_+)\,dx\ ,\ \theta ^2=\cosh (au_+-bv_+)\,dy\ ,\ \theta ^3=du\ ,\ \theta ^4=dv\ .\end{equation}On the half--null tetrad given via these 1--forms the 
 Ricci tensor components are 
 \begin{eqnarray}
 R_{ab}&=&-2ab\,\vartheta (u)\vartheta (v)\,g_{ab}-a\,\delta (u)(\tan bv_++\tanh bv_+)\delta ^3_a\delta ^3_b\nonumber\\
 &&-b\,\delta (v)(\tan au_++\tanh au_+)\delta ^4_a\delta ^4_b\ ,\label{3.10}\end{eqnarray}where $\delta (u)$ and $\delta (v)$ are the Dirac delta functions 
 singular on the null hyperplanes $u=0$ and $v=0$ respectively. The corresponding Newman--Penrose \cite{NP} components of the Weyl conformal curvature tensor 
 are  
 \begin{eqnarray}
 \Psi _0&=&\frac{1}{2}a\,\delta (u)\,(\tan bv_+-\tanh bv_+)+a^2\vartheta (u)\ ,\label{3.11}\\
 \Psi _1&=&0\ ,\label{3.12}\\
 \Psi _2&=&\frac{1}{3}\,ab\,\vartheta (u)\vartheta (v)\ ,\label{3.13}\\
 \Psi _3&=&0\ ,\label{3.14}\\
 \Psi _4&=&\frac{1}{2}b\,\delta (v)\,(\tan au_+-\tanh au_+)+b^2\vartheta (v)\ .\label{3.15}\end{eqnarray}

 \setcounter{equation}{0}
\section{The Nariai-Bertotti Space-Time}\indent
 
 In (\ref{3.10}) the delta function terms represent light--like shells \cite{BH} on $v>0, u=0$ (the coefficient of $a$) and on $u>0, v=0$ (the 
 coefficient of $b$) which form after the collision at $u=v=0$. The first term on the right hand side, which is non--zero for $u>0, v>0$, is the 
 cosmological constant term $\Lambda =-2a\,b$. We note that the Ricci tensor (\ref{3.10}) vanishes for $u<0$ and for $v<0$ so that these 
 pre--collision regions of space--time are vacuum regions (with the subregion $u<0$ \emph{and} $v<0$ Minkowskian space--time).
 
 In (\ref{3.11})--(\ref{3.15}) the delta function terms represent impulsive gravitational waves \cite{BH} on $u=0, v>0$ (the coefficient of $a$ in 
 (\ref{3.11})) and on $v=0, u>0$ (the coefficient of $b$ in (\ref{3.15})) which form after the collision. If $u<0$ then the only surviving component 
 of the Weyl tensor is $\Psi _4=b^2\vartheta (v)$. This is a vacuum region of space--time and $\Psi _4$ is the Riemann tensor of an incoming gravitational 
 shock wave. Similarly if $v<0$ then the only surviving component of the Riemann tensor is $\Psi _0=a^2\vartheta (u)$ which describes the second 
 incoming gravitational shock wave. When $u>0$ \emph{and} $v>0$ the only non--vanishing Weyl tensor components are $\Psi _0=a^2, 
 \Psi _2=\frac{1}{3}a\,b, \Psi _4=b^2$ and this is a Petrov type D Weyl tensor describing the Nariai--Bertotti homogeneous, vacuum gravitational field with 
 a cosmological constant. 
 
 It is remarkable that this head--on collision of gravitational shock waves generates a cosmological constant. It is well known that the energy in the incoming 
 waves can be distributed in different forms following the collisions of gravitational or electromagnetic shock waves. Following the head-on collision of 
 electromagnetic shock waves described by the Bell--Szekeres \cite{BS} solution, for example, two impulsive gravitational waves are generated traveling 
 away from each other (the delta function terms in (\ref{6.9}) and (\ref{6.13}) with $g_0=g_1=0$ below). Following the head--on collision of gravitational shock waves described above two impulsive 
 gravitational waves are also generated as well as two light--light shells of matter. The appearance of a cosmological constant can be interpreted as 
 part of a redistribution of the energy in the incoming waves since it represents energy in the form of a perfect fluid with proper density $\mu$ and 
 isotropic pressure $p$ satisfying the equation of state $\mu+p=0$ of dark energy. An alternative way of viewing this is that the system contains two vacua, 
 one with vanishing cosmological constant and one with a non--vanishing one, and the wave collision catalyzes a phase transition from one vacuum to another.
 
 \setcounter{equation}{0}
\section{In-Coming Waves}\indent
For convenience the space--time description of a light--like signal consisting of an electromagnetic shock wave accompanied by a gravitational shock wave can 
be written in a way that depends upon two parameters, a parameter $e_0$ associated with the electromagnetic shock wave and a parameter $g_0$ associated 
with the gravitational shock wave. If we wish to remove the electromagnetic shock wave we put $e_0=0$ and if we wish to remove the gravitational shock wave 
we put $g_0=0$. The expression for the metric tensor of the space-time depends upon the relative magnitudes of $e_0^2$ and $g_0^2$. If $e_0^2>g_0^2$ then
the line--element reads
\begin{equation}\label{5.1}
ds^2=-\cos^2\sqrt{e_0^2+g_0^2}\,u_+\,dx^2-\cos^2\sqrt{e_0^2-g_0^2}\,u_+\,dy^2+2du\,dv\ ,\end{equation}
where $u_+=u\,\vartheta (u)$ and again $\vartheta (u)$ is the Heaviside step function which equals unity if $u>0$ and vanishes if $u<0$. On the other hand if $e_0^2<g_0^2$ 
then the line--element reads
\begin{equation}\label{5.2}
ds^2=-\cos^2\sqrt{g_0^2+e_0^2}\,u_+\,dx^2-\cosh^2\sqrt{g_0^2-e_0^2}\,u_+\,dy^2+2du\,dv\ .\end{equation}We note the obvious fact that (\ref{5.1}) does not permit 
the special case $e_0=0$ while (\ref{5.2}) does not permit the special case $g_0=0$. Both of these line--elements take the form
\begin{equation}\label{5.3}
ds^2=-(\theta ^1)^2-(\theta ^2)^2+2\,\theta ^3\,\theta ^4=g_{ab}\theta ^a\,\theta ^b\ ,\end{equation}
with the 1--forms $\theta ^1=\cos\sqrt{e_0^2+g_0^2}\,u_+\,dx$, $\theta ^2=\cos\sqrt{e_0^2-g_0^2}\,u_+\,dy$, $\theta ^3=du$ and $\theta ^4=dv$ in the case of (\ref{5.1}), 
and the 1--forms $\theta ^1=\cos\sqrt{g_0^2+e_0^2}\,u_+\,dx$, $\theta ^2=\cosh\sqrt{g_0^2-e_0^2}\,u_+\,dy$, $\theta ^3=du$ and $\theta ^4=dv$ in the case of (\ref{5.2}), defining 
a half--null tetrad. The constants $g_{ab}$ are the tetrad components of the metric tensor and tetrad indices are raised and lowered using $g^{ab}$ (the components of the inverse 
of $g_{ab}$) and $g_{ab}$ respectively. We note that the hypersurfaces $u={\rm constant}$ and $v={\rm constant}$ are intersecting null hyperplanes in the space--times with line--elements (\ref{5.1}) and (\ref{5.2}). 
A calculation of the Ricci tensor components $R_{ab}$ on the half--null tetrad \emph{in either case} reveals that 
\begin{equation}\label{5.4}
R_{ab}=-2\,e_0^2\vartheta (u)\delta^3_a\delta^3_b\ .\end{equation}These are the vacuum Einstein--Maxwell field equations $R_{ab}=2\,E_{ab}$ with the electromagnetic energy tensor $E_{ab}=
F_{ac}\,F_b{}^c-\frac{1}{4}g_{ab}\,F_{cd}\,F^{cd}$ derived from the Maxwell 2--form
\begin{equation}\label{5.5}
F=\frac{1}{2}F_{ab}\,\theta ^a\wedge\theta ^b=e_0\vartheta (u)\,\theta ^1\wedge\theta ^3\ .\end{equation} It is easily checked that this is a solution of Maxwell's vacuum
field equations ($dF=0=d{}^*F$, where $d$ denotes exterior differentiation and the star denotes the Hodge dual). In addition (\ref{5.5}) is an algebraically special Maxwell 
field and is thus purely radiative with propagation direction in space--time that of the vector field $\partial /\partial v$. The wave profile is the step function and so we have here an 
electromagnetic shock wave. We also find that in the case of (\ref{5.1}) or (\ref{5.2}) the Newman--Penrose 
components $\Psi _A$ ($A=0, 1, 2, 3, 4$) of the Weyl conformal curvature tensor vanish except for 
\begin{equation}\label{5.6}
\Psi _0=g_0^2\vartheta (u)\ .\end{equation}This is a Petrov type N Weyl tensor with degenerate principle null direction $\partial /\partial v$ and it therefore represents a pure 
gravitational radiation field. The radiation has the step function profile and is therefore a shock wave.  This shock wave accompanies the electromagnetic shock wave (\ref{5.5}).

If we wish to consider similar light--like signals to those above but propagating in the opposite direction then labeling them with the pair of parameters $e_1, g_1$ the corresponding 
line--elements are: if $e_1^2>g_1^2$ then
\begin{equation}\label{5.7}
ds^2=-\cos^2\sqrt{e_1^2+g_1^2}\,v_+\,dx^2-\cos^2\sqrt{e_1^2-g_1^2}\,v_+\,dy^2+2du\,dv\ ,\end{equation}and if $e_1^2<g_1^2$ then
\begin{equation}\label{5.8}
ds^2=-\cos^2\sqrt{g_1^2+e_1^2}\,v_+\,dx^2-\cosh^2\sqrt{g_1^2-e_1^2}\,v_+\,dy^2+2du\,dv\ ,\end{equation}with $v_+=v\,\vartheta (v)$. In principle one should be able to 
obtain the space--time following the collision (at $u=v=0$) of either of (\ref{5.1}) or (\ref{5.2}) with either of (\ref{5.7}) or (\ref{5.8}). This remains an open question. In the next section 
we give the space--time following the collision of the signal described by (\ref{5.1}) with the signal described by (\ref{5.7}) and the space--time following the collision of 
the signal described by (\ref{5.2}) with the signal described by (\ref{5.8}), and this only if the simplifying assumption
\begin{equation}\label{5.9}
e_0g_1=e_1g_0\ ,\end{equation}is satisfied by the parameters involved. In both of these cases the space--time in the interaction region following the collision is a Bertotti \cite{B} space--time.

\setcounter{equation}{0}
\section{Post Collision Space-Times}\indent
If $e_0^2>g_0^2$ and (\ref{5.9}) is satisfied (and thus $e_1^2>g_1^2$) then the solution of the collision problem we propose is described by the line--element
\begin{equation}\label{6.1}
ds^2=-\cos^2\Psi\,dx^2-\cos^2\phi\,dy^2+2\,du\,dv\ ,\end{equation}
where
\begin{eqnarray}
\Psi&=&\sqrt{e_0^2+g_0^2}\,u_++\sqrt{e_1^2+g_1^2}\,v_+\ ,\label{6.2}\\
\phi&=&\sqrt{e_0^2-g_0^2}\,u_+-\sqrt{e_1^2-g_1^2}\,v_+\ .\label{6.3}\end{eqnarray}If $e_0^2<g_0^2$ and (\ref{5.9}) is satisfied (and thus $e_1^2<g_1^2$) the 
solution we propose is given by the line--element 
\begin{equation}\label{6.4}
ds^2=-\cos^2\Psi\,dx^2-\cosh^2\chi\,dy^2+2\,du\,dv\ ,\end{equation}
where $\Psi$ is given in (\ref{6.2}) and 
\begin{equation}\label{6.5}
\chi=\sqrt{g_0^2-e_0^2}\,u_+-\sqrt{g_1^2-e_1^2}\,v_+\ .\end{equation}In the case of (\ref{6.1}) or (\ref{6.4}) the line--element can again be written in the form (\ref{5.3}) 
with now the 1--forms given either by $\theta ^1=\cos\Psi\,dx$, $\theta ^2=\cos\phi\,dy$, $\theta ^3=du$ and $\theta ^4=dv$ or by 
$\theta ^1=\cos\Psi\,dx$, $\theta ^2=\cosh\chi\,dy$, $\theta ^3=du$ and $\theta ^4=dv$ respectively. In the case of (\ref{6.1}) the Ricci tensor components on the half--null 
tetrad given via the 1--forms can be written in the form
\begin{eqnarray}
R_{ab}&=&\Lambda\,\vartheta(u)\,\vartheta(v)\,g_{ab}+2\,E_{ab}+\biggl\{\sqrt{e_0^2-g_0^2}\,\tan\sqrt{e_1^2-g_1^2}\,v_+\nonumber\\
&&-\sqrt{e_0^2+g_0^2}\,\tan\sqrt{e_1^2+g_1^2}\,v_+\biggr\}\,\delta (u)\,\delta ^3_a\delta ^3_b+\biggl\{\sqrt{e_1^2-g_1^2}\,\tan\sqrt{e_0^2-g_0^2}\,u_+
\nonumber\\&&-\sqrt{e_1^2+g_1^2}\,\tan\sqrt{e_0^2+g_0^2}\,u_+\biggr\}\,\delta (v)\,\delta ^4_a\delta ^4_b\ ,\label{6.6}\end{eqnarray}
where
\begin{equation}\label{6.7}
\Lambda =-2\,g_0g_1\ ,\end{equation}and the components $E_{ab}=E_{ba}$ are identically zero except for
\begin{equation}\label{6.8}
E_{11}=-E_{22}=e_0e_1\vartheta (u)\,\vartheta (v)\ ,\ \ E_{33}=-e_0^2\vartheta (u)\ ,\ \ E_{44}=-e_1^2\vartheta (v)\ .\end{equation}
The Newman--Penrose components of the Weyl conformal curvature tensor for (\ref{6.1}) are
\begin{eqnarray}
\Psi_0&=&\frac{1}{2}\biggl\{\sqrt{e_0^2-g_0^2}\,\tan\sqrt{e_1^2-g_1^2}\,v_+\nonumber\\
&+&\sqrt{e_0^2+g_0^2}\,\tan\sqrt{e_1^2+g_1^2}\,v_+\biggr\}\,\delta (u)+g_0^2\vartheta (u)\ ,\label{6.9}\\
\Psi_1&=&0\ ,\label{6.10}\\
\Psi_2&=&\frac{1}{3}\,g_0\,g_1\,\vartheta (u)\,\vartheta (v)\ ,\label{6.11}\\
\Psi_3&=&0\ ,\label{6.12}\\
\Psi_4&=&\frac{1}{2}\biggl\{\sqrt{e_1^2+g_1^2}\,\tan\sqrt{e_0^2+g_0^2}\,u_+\nonumber\\
&+&\sqrt{e_1^2-g_1^2}\,\tan\sqrt{e_0^2-g_0^2}\,u_+\biggr\}\,\delta (v)+g_1^2\vartheta (v)\ . \label{6.13}\end{eqnarray}
For (\ref{6.4}) the Ricci tensor components on the half--null tetrad are given by
\begin{eqnarray}
R_{ab}&=&\Lambda\,\vartheta(u)\,\vartheta(v)\,g_{ab}+2\,E_{ab}-\biggl\{\sqrt{g_0^2+e_0^2}\,\tan\sqrt{g_1^2+e_1^2}\,v_+\nonumber\\
&&+\sqrt{g_0^2-e_0^2}\,\tanh\sqrt{g_1^2-e_1^2}\,v_+\biggr\}\,\delta (u)\,\delta ^3_a\delta ^3_b-\biggl\{\sqrt{g_1^2+e_1^2}\,\tan\sqrt{g_0^2+e_0^2}\,u_+
\nonumber\\&&+\sqrt{g_1^2-e_1^2}\,\tanh\sqrt{g_0^2-e_0^2}\,u_+\biggr\}\,\delta (v)\,\delta ^4_a\delta ^4_b\ ,\label{6.14}\end{eqnarray}
with $\Lambda$ and $E_{ab}$ as in (\ref{6.7}) and (\ref{6.8}) and the Newman--Penrose components of the Weyl conformal curvature tensor 
are 
\begin{eqnarray}
\Psi_0&=&\frac{1}{2}\biggl\{\sqrt{g_0^2+e_0^2}\,\tan\sqrt{g_1^2+e_1^2}\,v_+\nonumber\\
&-&\sqrt{g_0^2-e_0^2}\,\tanh\sqrt{g_1^2-e_1^2}\,v_+\biggr\}\,\delta (u)+g_0^2\vartheta (u)\ ,\label{6.15}\\
\Psi_1&=&0\ ,\label{6.16}\\
\Psi_2&=&\frac{1}{3}\,g_0\,g_1\,\vartheta (u)\,\vartheta (v)\ ,\label{6.17}\\
\Psi_3&=&0\ ,\label{6.18}\\
\Psi_4&=&\frac{1}{2}\biggl\{\sqrt{g_1^2+e_1^2}\,\tan\sqrt{g_0^2+e_0^2}\,u_+\nonumber\\
&-&\sqrt{g_1^2-e_1^2}\,\tanh\sqrt{g_0^2-e_0^2}\,u_+\biggr\}\,\delta (v)+g_1^2\vartheta (v)\ . \label{6.19}\end{eqnarray}
In the case of (\ref{6.1}) and (\ref{6.4}), $E_{ab}$ given by (\ref{6.8}) is the electromagnetic energy tensor of a Maxwell 2--form
\begin{equation}\label{6.20}
F=e_0\vartheta (u)\,\theta ^1\wedge\theta ^3+e_1\vartheta (v)\,\theta ^1\wedge\theta ^4\ .\end{equation}
It is easily checked that on account of (\ref{5.9}) this satisfies Maxwell's vacuum field equations in the space--times with line--elements 
(\ref{6.1}) and (\ref{6.4}) with the appropriate choice of 1--forms in each case (indicated following (\ref{6.5}) above).

 \setcounter{equation}{0}
\section{The Bertotti Space--Time}\indent
In the space--times with line--elements (\ref{6.1}) and (\ref{6.4}) the (overlapping) regions prior to the collision of the light--like signals 
correspond to $u<0$ with line--element (\ref{5.7}) or (\ref{5.8}) and $v<0$ with line--element (\ref{5.1}) or (\ref{5.2}). The post--collision region 
of the space--times with line--elements (\ref{6.1}) and (\ref{6.4}) is $u>0, v>0$. We see from the Ricci tensor components (\ref{6.6}) or (\ref{6.14}) 
that the boundaries of this post--collision region, $v=0$ with $u>0$ and $u=0$ with $v>0$, are the histories of plane, light--like shells of matter (e.g. bursts 
of neutrinos) \cite{BH} traveling away from each other with the speed of light in the directions of the incoming signals. These objects are described by the delta function terms in (\ref{6.6}) 
and (\ref{6.14}). The delta function terms in the Weyl tensor components (\ref{6.9}) and (\ref{6.13}) or (\ref{6.15}) and (\ref{6.19}) describe 
plane, impulsive gravitational waves \cite{BH} accompanying these light--like shells. The Maxwell 2--form (\ref{6.20}) describes the two incoming 
electromagnetic shock waves and also gives the resulting electromagnetic field in the region $u>0, v>0$. We see that in this post--collision region 
of space--time (\ref{6.6}) and (\ref{6.14}) both simplify to 
\begin{equation}\label{7.1}
R_{ab}=\Lambda\,g_{ab}+2\,E_{ab}\ ,\end{equation}with $\Lambda$ given by (\ref{6.7}) and $E_{ab}$ by (\ref{6.8}) when $u>0$ and $v>0$. Thus 
$\Lambda$ is a cosmological constant. This homogeneous space--time with line--element (\ref{6.1}) or (\ref{6.4}) with $u>0, v>0$ is thus a solution 
of the Einstein--Maxwell vacuum field equations with a cosmological constant and is a Bertotti \cite{B} space--time. Finally we note that (\ref{6.1}) specialized 
by taking $g_0=0=g_1$ yields the Bell--Szekeres \cite{BS} line--element while (\ref{6.4}) specialized by taking $e_0=0=e_1$ yields the line--element 
given in (\ref{3.8}) and (\ref{3.9}).

An alternative interpretation of the post collision region, mentioned at the end of section 3, is that dark energy in the form of a perfect fluid with isotropic tension equal to proper energy density 
is present. Thus the collisions described in this paper could be interpreted as a source of this special kind of matter.

\end{document}